\documentclass[twocolumn]{aastex701}
\usepackage{float}

\usepackage{amsfonts}
\usepackage{booktabs}
\let\tablenum\relax
\usepackage{siunitx}
\DeclareSIUnit\bar{bar}
\usepackage{xcolor}
\usepackage{url, microtype, isomath, mathtools, siunitx, physics}
\usepackage{graphicx}
\usepackage{rotating}
\usepackage{booktabs}
\usepackage[version=4]{mhchem}
\usepackage{multirow}
\usepackage{rotfloat}
\usepackage{rotating}
\usepackage{mathrsfs}

\begin{document}

\title{Improved Constraints on the Surface of LHS 3844 b from its Mid-Infrared Spectrum}

\author[orcid=0000-0003-0062-1168]{Kimberly Paragas}
\affiliation{Division of Geological and Planetary Sciences, California Institute of Technology}
\email[show]{kparagas@caltech.edu}  

\author[orcid=0000-0002-5375-4725]{Heather A. Knutson} 
\affiliation{Division of Geological and Planetary Sciences, California Institute of Technology}
\email{heather@caltech.edu}

\author[orcid=0000-0003-2215-8485]{Renyu Hu} 
\affiliation{Department of Physics and Astronomy, Rutgers, The State University of New Jersey, Piscataway, NJ 08854, USA}
\affiliation{Department of Astronomy \& Astrophysics, The Pennsylvania State University, University Park, PA 16802, USA}
\affiliation{Center for Exoplanets and Habitable Worlds, The Pennsylvania State University, University Park, PA 16802, USA}
\affiliation{Institute for Computational and Data Science, The Pennsylvania State University, University Park, PA 16802, USA}
\affiliation{Jet Propulsion Laboratory, California Institute of Technology, Pasadena, CA 91109, USA}
\email{rh1071@physics.rutgers.edu}

\author[orcid=0000-0002-2745-3240]{Bethany L. Ehlmann} 
\affiliation{Division of Geological and Planetary Sciences, California Institute of Technology}
\affiliation{Laboratory for Atmospheric \& Space Physics, University of Colorado Boulder, CO 80309, USA}
\affiliation{Department of Earth Science, University of Colorado Boulder, CO 80309, USA}
\affiliation{Department of Astrophysical \& Planetary Sciences, University of Colorado Boulder, CO 80309, USA}
\email{Bethany.Ehlmann@lasp.colorado.edu}

\author[orcid=0000-0003-0971-1709]{Aishwarya R. Iyer}
\affiliation{NASA Goddard Space Flight Center, Greenbelt, MD 20771, USA}
\affiliation{NASA Postdoctoral Fellowship through Oak Ridge Associated Universities}
\email{aishwarya.iyer@nasa.gov}

\author[orcid=0000-0003-0562-6750]{Sebastian Zieba} 
\affiliation{Center for Astrophysics, Harvard \& Smithsonian, Cambridge, MA 02138, USA}
\email{sebastian.zieba@cfa.harvard.edu}

\author[orcid=0000-0003-0514-1147]{Laura Kreidberg} 
\affiliation{Max-Planck-Institut f\"ur Astronomie, D-69117 Heidelberg, Germany}
\email{kreidberg@mpia.de}

\begin{abstract}
 The thermal emission spectra of terrestrial exoplanets provide a window into their surface compositions and corresponding geological processes. Close-in, rocky planets orbiting M dwarfs are ideal targets for these studies, as observations have shown that most have little to no atmosphere. LHS 3844 b is an ultra-short period super-Earth orbiting a nearby M dwarf, and is the most favorable target that does not have a dayside magma ocean. We present an updated JWST MIRI/LRS emission spectrum ($5-12$~\unit{\micro\meter}) of this planet, derived from eight new and three archival eclipse observations. We combine these visits to obtain the highest SNR emission spectrum for any rocky exoplanet thus far, with a median SNR of 29 across 12 wavelength bins. Our data are sensitive to the predicted Si-O stretching feature (the strongest mid-infrared silicate feature) for a wide range of common materials and the transparency feature, which is relatively insensitive to composition and constrains the grain size and porosity of surface materials. We find that LHS 3844~b's emission spectrum resembles a blackbody shortward of \SI{10}{\micro\meter}, disfavoring the presence of a transparency feature, and identify a tentative feature ($1.9-3.5\sigma$) between $10-12$~\unit{\micro\meter}. If this feature is astrophysical and not instrumental in nature, it can be matched by a fayalite (the iron end-member of olivine) surface, but might also be consistent with other candidate planet-forming materials like spinel or silicon carbide. We discuss possible biases introduced by the wavelength-dependent instrumental ramp, and outline several strategies to mitigate this effect in future observations of LHS 3844~b. 

\end{abstract}

\section{Introduction} 
Enabled by the unprecedented sensitivity of JWST, the search for atmospheres on rocky exoplanets orbiting M dwarfs has become one of the leading pursuits in exoplanetary science \citep{Kreidberg2025}. Over the past five years, a wide range of JWST programs have searched for evidence of atmospheric features in the transmission and emission spectra of these planets, including the ongoing 500-hour Rocky Worlds Director's Discretionary Time (DDT) program. These surveys indicate that most of these rocky planets have little to no atmosphere \citep{Kreidberg2019, Crossfield2022, Greene2023, Zieba2023, WeinerMansfield2024, Xue2024, Zhang2024, Allen2025, Fortune2025, Luque2025, MeierValdes2025, Wachiraphan2025, Xue2025, Connors2026, Gillon2026, Holmberg2026, Rochon2026}, aside from the subset that are able to maintain atmospheres via sustained outgassing from a molten dayside or tidally driven volcanism \citep[e.g.,][]{Hu2024, Bello-Arufe2025, Monaghan2025, Teske2025, Coy2026}. 

The airless rocky planets present a valuable opportunity for surface characterization via thermal emission measurements in the near- (NIR) and mid-infrared (MIR). Different surfaces will exhibit distinct spectral features in this wavelength range that can be used to constrain their compositions \citep[e.g.,][]{Hu2012, Whittaker2022, Hammond2025, Paragas2025}. For example, the surface \ce{SiO2} content can be inferred by measuring the location of the MIR Christiansen feature \citep{Conel1969} and the corresponding Si-O stretching feature \citep{Salisbury1993}. The surface compositions of these planets in turn constrain their interior dynamics, including the depth and vigor of mantle convection, which has implications for crustal recycling rates  \citep[see review by ][and references therein]{Guimond2024}. In a separate dimension of characterization, older planetary surfaces will be particulate in nature, due to the effects of ongoing meteorite impacts, while newer surfaces will contain much larger slabs of solid, rocky material. These two surface types can be differentiated by the strength of the transparency feature, which is created by multiple scattering in powdered materials and single scattering in rock slabs \citep[e.g.,][]{Paragas2025}. At the same time, the retrieved dayside temperature can be used to infer the planet's surface albedo, which is sensitive to the composition, particle size, and amount of space weathering \citep[e.g.,][]{Lyu2024, Coy2025}. This means that emission spectroscopy of airless rocky exoplanets with solid (i.e., not molten) dayside surfaces offers a rich window into their surface properties and corresponding geological histories.

Among this set of planets, LHS 3844 b stands out as the most favorable target for surface characterization. With a radius of $R_p=1.303\pm0.022$ R$_\oplus$ and a mass of $M_p=2.27\pm0.23$ M$_\oplus$ \citep{Vanderspek2019, Hacker2026, Nagel2026}, its average density is consistent with an Earth-like bulk composition. Although it has an orbital period of just $11.1$~\unit{\hour}, it still has a relatively low equilibrium temperature of $T_\mathrm{eq}=816$~\unit{\kelvin} thanks to the small radius ($R_s = 0.189$ R$_\odot$) and low effective temperature ($T_\mathrm{eff}=3080$~\unit{\kelvin}) of its mid-M dwarf host \citep{Vanderspek2019, Nagel2026}. This low equilibrium temperature corresponds to a maximum predicted dayside temperature of 1045~\unit{\kelvin} at the substellar point for the zero albedo case, below the melting point of most silicates. 

Published observations of LHS 3844~b place strict upper limits on the thickness of its atmosphere, indicating that its thermal emission spectrum should be dominated by surface features. \citet{Kreidberg2019} used Spitzer to measure LHS 3844~b's infrared phase curve in the \SI{4.5}{\micro\meter} bandpass and concluded that its large day-night brightness temperature difference is best matched by atmosphere models with surface pressures less than 0.1 bars. Follow-up ground-based transmission spectroscopy of the planet was consistent with a flat line and ruled out atmospheric scenarios with pressures greater than 0.1 bars \citep{Diamond-Lowe2020}. More recently, \citet{Zieba2026} measured its mid-infrared dayside emission spectrum with MIRI LRS and found that it is consistent with a blackbody, with no detected spectral features. This study placed even tighter upper limits of \SI{100}{\milli\bar} ($5\sigma$) and \SI{10}{\micro\bar} ($3\sigma$) surface pressures for atmospheres with trace concentrations of \ce{CO2} and \ce{SO2}, respectively. Although these observations also placed tight constraints on the planet's dayside Bond albedo, they did not detect any statistically significant surface spectral features and therefore could be matched by a broad range of surface compositions \citep{Zieba2026}. 

Here we present an improved JWST MIRI LRS emission spectrum for LHS 3844 b constructed from a total of 11 secondary eclipses, including the three published eclipses from \citet{Zieba2026} and eight new eclipses from GO 7953 (PI: K. Paragas). In Section~\ref{sec:methods} we describe our analysis of the JWST observations including both the planetary and stellar mid-IR emission spectra. In Section~\ref{sec:blackbody}, we fit LHS 3844~b's measured emission spectrum with a blackbody, report a tentative detection of a feature centered at \SI{11.6}{\micro\meter}, and consider whether the feature might be caused by instrumental effects. In Section~\ref{sec:surface_composition}, we present surface spectral models motivated by the tentative feature, show that the measured spectrum disfavors fine-grained powder materials, and place albedo constraints based on surface material. In Section~\ref{sec:discussion_and_conclusions}, we discuss the implications of our results for its surface properties as well as provide recommendations for future observational approaches that can improve the quality of the derived spectra and enable refined interpretations of the planet's surface composition.

\section{Methods} \label{sec:methods}
\subsection{Observations}
We analyzed 11 secondary eclipse observations of LHS 3844 b taken with the MIRI instrument in the low-resolution spectroscopy (slitless) mode onboard JWST. These observations include three eclipses from Program GO 1846 that were previously published in \cite{Zieba2026} and 8 new eclipses from Program GO 7953. More information on each individual observation can be found in Table~\ref{tab:main}. Observations from both programs were obtained using 30 groups per integration, but we increased the overall observing time in the new visits from 2.58 hours (1887 integrations) to 2.84 hours (2225 integrations) in order to allow for better modeling of the exponential ramp at the beginning of each observation. See \citet{Pontoppidan2016} for detailed explanations of the exposure configuration.

\subsection{Data Reduction with \texttt{Eureka!}}
We used \texttt{Eureka!} \citep{Bell2022} version 1.3 with \texttt{jwst} pipeline version 1.18.0 and \texttt{crds} version 13.0.6. for Stages 1 through 4, which include detector processing, data calibration, data reduction, and light curve generation, respectively. For Stage 1, we used the default settings except for \texttt{jump\_rejection\_threshold}, \texttt{skip\_emicorr}, and \texttt{skip\_rcsd}. As recommended for TSO data, we increased the \texttt{jump\_rejection\_threshold} to 7. We tested True and False settings of both \texttt{skip\_emicorr}, a step that corrects for electromagnetic interference (EMI) noise in the raw data when set to False, and \texttt{skip\_rcsd}, a step that flags groups that are affected by nonlinear transients that may occur from exponential settling of the detector when set to False. We kept the configuration settings that minimized the median absolute deviation (MAD) in the white light curves. All visits preferred \texttt{skip\_rcsd} = True. For Stage 2, we set \texttt{skip\_photom} to True as recommended by \texttt{Eureka!} and used the default settings for the remaining parameters. For Stage 3, we used the same values for every MIRI LRS eclipse. We defined \texttt{ywindow} and \texttt{xwindow} as [109, 388] and [11, 62], respectively. We set \texttt{ff\_outlier} to True and \texttt{bg\_thresh} to [np.inf, np.inf]. We tested a number of combinations for the background and spectral apertures (\texttt{bg\_hw} and \texttt{spec\_hw}, respectively), and selected the combination that minimized the MAD in the white light curve residuals. The values for the other background and spectral extraction parameters can be found in Table~\ref{tab:main}. For Stage 4, we generated white light curves that span \SIrange{5.06}{12.368}{\micro\meter}, and divided each into 12 spectroscopic channels to generate the spectroscopic light curves. We followed the same approach as \citet{Zieba2026} for removing outliers by clipping binned data that deviate by more than $4\sigma$ using a 20 integration wide box-car filter. We discarded less than 0.5\% of integrations for each visit.

\startlongtable
\begin{deluxetable*}{lcccccccl}
\tabletypesize{\footnotesize}
\tablecaption{Summary of JWST observations of LHS~3844b analyzed in this work. \label{tab:main}}
\tablehead{
\colhead{Visit} & \colhead{JWST Obs. \#} & \colhead{Date (UT)} & \colhead{Start (UT)} & \colhead{End (UT)} & \colhead{skip\_emicorr} & \colhead{spec\_hw} & \colhead{bg\_hw} & \colhead{Fit Parameters}
}
\startdata
GO 1846 \#1 & 1 & Jun. 19 2023 & 02:52:07 & 06:06:27 & False & 3 & 10 & $c_0$, $c_1$, $r_0$, $r_1$, \\
 & & & & & & & & $c_\mathrm{xpos}$, $c_\mathrm{ypos}$ \\
\tableline
GO 1846 \#2 & 102 & Sep. 30, 2023 & 08:15:59 & 11:55:36 & False & 3 & 12 & $c_0$, $r_0$, $r_1$, \\
 & & & & & & & & $c_\mathrm{ypos}$, $c_\mathrm{ywidth}$ \\
\tableline
GO 1846 \#3 & 3 & May 3 2024 & 23:04:50 & 03:06:08 & False & 2 & 16 & $c_0$, $r_0$, $r_1$, $c_\mathrm{xpos}$, \\
 & & & & & & & & $c_\mathrm{ypos}$, $c_\mathrm{xwidth}$, $c_\mathrm{ywidth}$ \\
\tableline
GO 7953 \#1 & 6 & Jul. 10 2025 & 07:22:35 & 11:59:55 & True & 5 & 8 & $c_0$, $c_1$, $r_0$, $r_1$, \\
 & & & & & & & & $c_\mathrm{ypos}$, $c_\mathrm{xwidth}$, $c_\mathrm{ywidth}$ \\
\tableline
GO 7953 \#2 & 3 & Aug. 16 2025 & 08:12:10 & 12:49:23 & False & 5 & 8 & $c_0$, $r_0$, $r_1$, \\
 & & & & & & & & $c_\mathrm{ypos}$, $c_\mathrm{ywidth}$ \\
\tableline
GO 7953 \#3 & 2 & Sep. 29 2025 & 19:43:48 & 23:41:31 & False & 4 & 8 & $c_0$, $r_0$, $r_1$, $c_\mathrm{xpos}$, \\
 & & & & & & & & $c_\mathrm{ypos}$, $c_\mathrm{xwidth}$, $c_\mathrm{ywidth}$ \\
\tableline
GO 7953 \#4 & 4 & Sep. 30 2025 & 06:29:41 & 10:33:48 & False & 3 & 12 & $c_0$, $r_0$, $r_1$, \\
 & & & & & & & & $c_\mathrm{ypos}$, $c_\mathrm{xwidth}$ \\
\tableline
GO 7953 \#5 & 5 & Oct. 1 2025 & 16:01:40 & 19:53:43 & False & 2 & 6 & $c_0$, $c_1$, $r_0$, $r_1$, \\
 & & & & & & & & $c_\mathrm{xpos}$ \\
\tableline
GO 7953 \#6 & 7 & Oct. 2 2025 & 02:41:24 & 07:00:31 & False & 3 & 20 & $c_0$, $r_0$, $r_1$, \\
 & & & & & & & & $c_\mathrm{xpos}$, $c_\mathrm{ypos}$, $c_\mathrm{xwidth}$ \\
\tableline
GO 7953 \#7 & 8 & Oct. 3 2025 & 00:58:42 & 05:13:55 & False & 6 & 6 & $c_0$, $c_1$, $r_0$, $r_1$, \\
 & & & & & & & & $c_\mathrm{ypos}$, $c_\mathrm{xwidth}$, $c_\mathrm{ywidth}$ \\
\tableline
GO 7953 \#8 & 1 & Oct. 16 2025 & 23:42:55 & 04:20:11 & False & 4 & 12 & $c_0$, $c_1$, $r_0$, $r_1$, \\
 & & & & & & & & $c_\mathrm{xpos}$, $c_\mathrm{ypos}$, $c_\mathrm{xwidth}$ \\
\enddata
\tablecomments{The fit parameters are described in the Light Curve Modeling subsection in Section~\ref{sec:methods}.}
\end{deluxetable*}

\subsection{Light Curve Modeling}
\label{sec:lc_modeling}
We used Markov Chain Monte Carlo (MCMC) with \texttt{emcee} \citep{Foreman-Mackey2013} to fit the white light curves for each secondary eclipse with a \texttt{batman} \citep{Kreidberg2015} eclipse model combined with a sinusoidal phase curve model and a systematics model. For the eclipse model, we fixed the orbital period $P = 0.462929709$~\unit{\day}, inclination $i=89^\circ$, the ratio of the planetary semi-major axis to the stellar radius $a/R_* = 7.122$, and the planet-star radius ratio $R_p/R_* = 0.06262$ to the values reported in \citet{Nagel2026}. We assumed a circular orbit and allowed the eclipse center time $t_0$ and depth $f_p/f_*$ to vary as free parameters for each individual eclipse observation. For the phase curve model, we use a first order cosine (simple) model $F_{pc, 1} = 1 + \frac{A_1}{2}(cos{\phi}-1)$, where $F_{pc}$ is the phase curve flux, $A_1$ is the phase curve amplitude, and $\phi$ is given by $\phi = 2\pi (t-t_{sec})/P$ with $t_{sec}$ being the secondary eclipse time. We found that the data do not meaningfully constrain the phase curve amplitude $A_1$ in our individual eclipse fits. We therefore chose to fix it to 1 (corresponding to the case of zero emission from the planet's night side, as expected for an airless planet with no internal heat sources), similar to \citet{Zieba2026}. We modeled scenarios where the nightside flux was nonzero, and even with a nightside temperature of \SI{500}{\kelvin}, the largest difference from the zero nightside flux case is 12~ppm, less than what could be detectable with our data, over the phase range of our observations. Additionally, we fit the data assuming a first and second order cosine phase curve model using $F_{pc, 2} = 1 + \frac{A_1}{2}(cos{\phi}-1) + \frac{A_2}{2} (cos(2\phi) - 1)$, where $A_2$ is the amplitude of a secondary cosine term, and we fixed $A_1$ and $A_2$ to best-fit values from the NIRSpec phase curve of the planet (S. Zieba et al. in preparation). The resulting averaged spectrum is consistent with the simple phase curve model $F_{pc,1} (A_1 = 1)$ case, but it is worth noting that the eclipse depths are consistently shallower by about 10~ppm until \SI{10}{\micro\meter} and then deeper at the two longest wavelength bins at \SI{11.5}{\micro\meter} and \SI{12}{\micro\meter} compared to the averaged spectrum assuming the simple phase curve model. 

\begin{figure*}[ht!]
    \centering
    \includegraphics[width=0.97\textwidth]{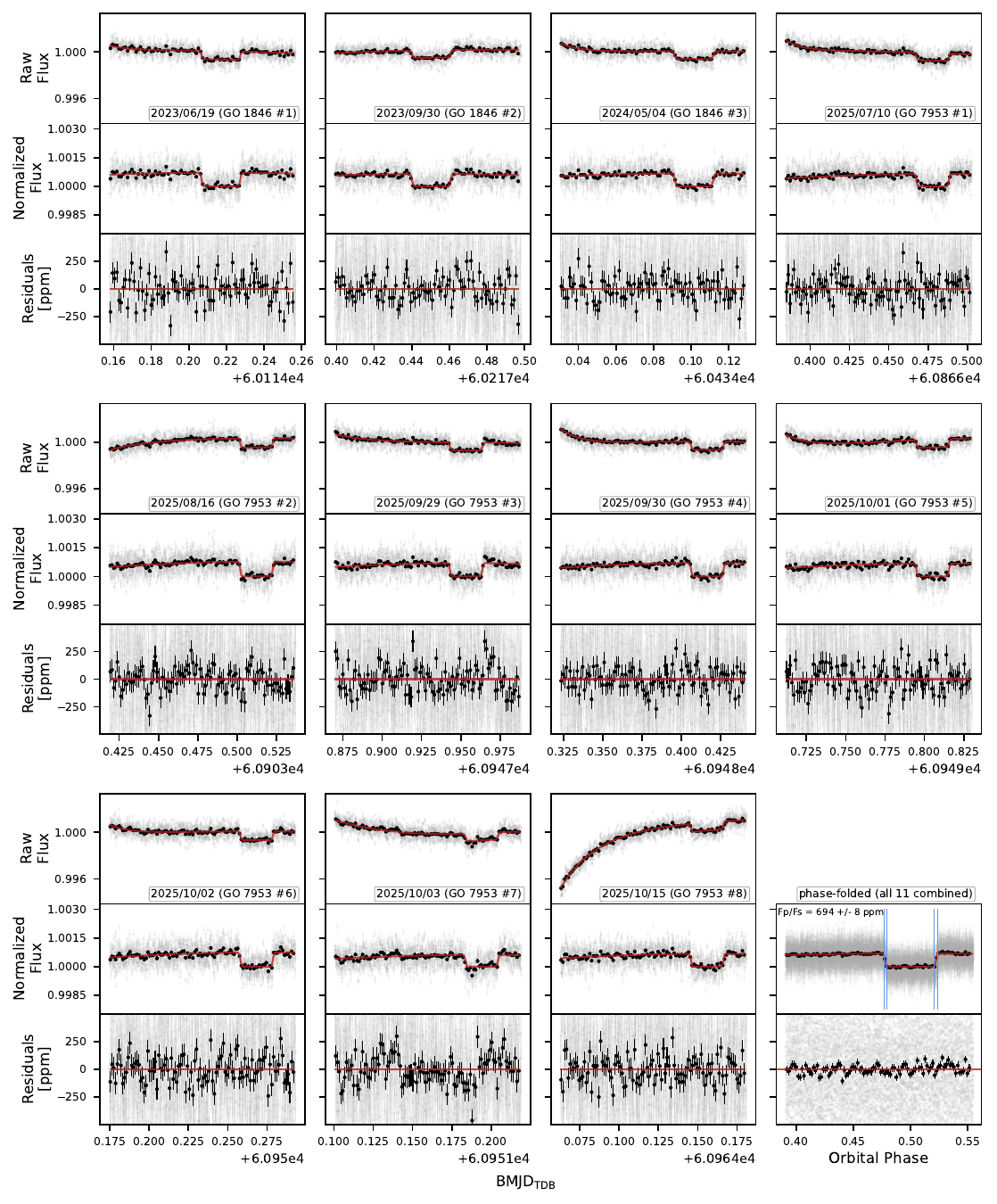}
    \caption{The 11 three-panel plots show the raw flux, normalized flux, and residuals of the white light curves for each MIRI LRS eclipse with their best-fit model overplotted in red. The two-panel figure in the bottom right is the phase-folded combined white light curve of all 11 visits with the blue vertical lines indicating T1-T4, i.e., ingress and egress. The gray points are the raw data at the native resolution of the observations while the black points are the binned data to a cadence of 1.6 minutes.}
    \label{fig:white_light_curves}
\end{figure*}

Our initial systematics model for each eclipse included a linear trend ($c_0$ and $c_1$ as the constant and slope, respectively), an exponential ramp ($r_0$ and $r_1$ as the amplitude and decay constant, respectively), and linear coefficients for the measured pointing drift and changes in the PSF width in both the x and y directions, the dispersion and spatial directions, ($c_\mathrm{xpos}$, $c_\mathrm{ypos}$, $c_\mathrm{xwidth}$, and $c_\mathrm{ywidth}$). For each observation, we decided which components, except for the constant trend $c_0$ which is used universally, of this instrumental model to include by minimizing the Bayesian Information Criterion \citep[BIC;][]{Schwarz1978}. For cases where the difference between the best and second best models was a statistical wash ($\Delta\mathrm{BIC} < 3$), we picked the model with fewer free parameters. The resulting white light curves and best-fit model are shown in Figure~\ref{fig:white_light_curves}. The fit parameters used for each observation are listed in Table~\ref{tab:main}. 

To assess the quality of our white light curve fits, we calculated the root mean square (rms) of the residuals as a function of bin size from 0 (unbinned data) to 207 bins ($\sim$~10\% of the data in each bin) to create time averaging plots and compared this to the expected noise scaling of $\sqrt n_{bins}$ \citep{Pont2006}. We found that most observations were well behaved, with minimal evidence for correlated noise. However, there were two observations that had some excess noise on longer timescales. The observation taken on September 29 2025 has a few upticks in flux that occurred 12, 84, and 149 minutes into our observation. The last uptick coincided with the end of egress and may be a flare. The observation taken on October 3 2025 has a residual structure that spans 60951.141 to 60951.192 BMJD$_\mathrm{TDB}$ with consistently low flux values. Although a qualitatively similar structure appeared in the y centroid timeseries, this feature did not vanish when we included this parameter as a detrending vector in our fit. Despite these issues, we found that the dayside emission spectra from these two visits were in good agreement with the other nine visits, and therefore included them in our averaged dayside emission spectrum.

\begin{figure}
    \centering
    \includegraphics[width=\linewidth]{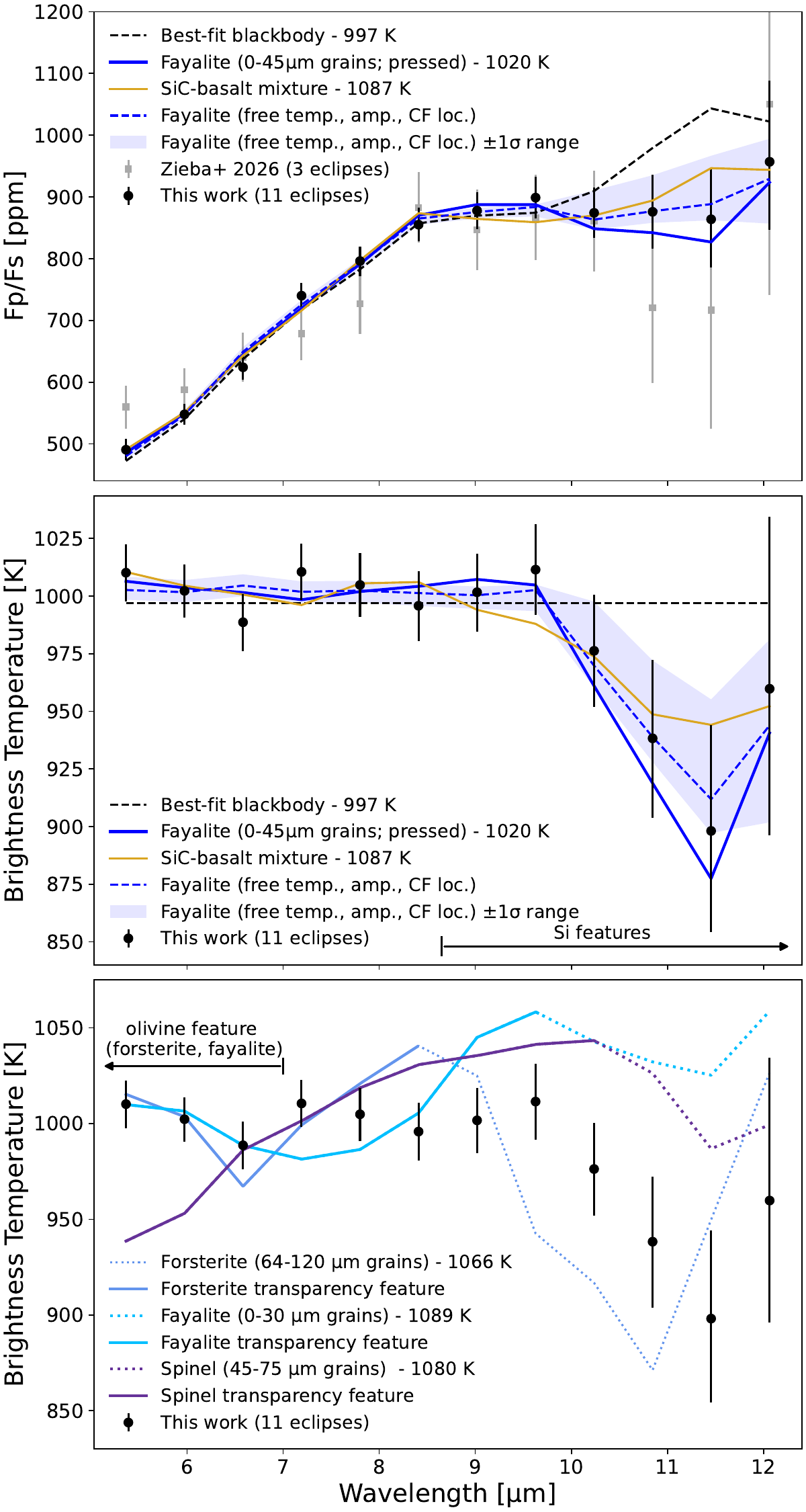}
    \caption{Top: Planet-to-star flux ratio emission spectrum from the three eclipses published in \citet{Zieba2026}, compared to the combined spectrum from all 11 eclipses in this work (including the three previously published). Middle and bottom: The same spectrum converted to brightness temperature to highlight the feature near 10--12~\unit{\micro\meter}. The top two panels show the best-fit blackbody, fayalite, and SiC-basalt mixture models (each with temperature as a free parameter), and the three parameter Fayalite model with temperature (temp.), amplitude (amp.), 
    and Christiansen Feature (CF) location (loc.) as free parameters. The bottom panel shows the best-fit forsterite, fayalite, and spinel models of small, loosely packed grains with their corresponding transparency features denoted by the solid lines.}
    \label{fig:emission_spectrum}
\end{figure}

For the spectroscopic light curve fits, we followed \citet{Zieba2026} by binning the data into 12 channels between \SIrange{5.06}{12.368}{\micro\meter}, each with a width of \SI{0.609}{\micro\meter}. We fit the resulting light curves in each channel with the same set of parameters as the white light curves except for the transit time $t_0$, which we fixed to the best-fit value from the white light curve fit. We then took the inverse-variance weighted average of the 11 individual eclipse depths in each bandpass to construct the averaged emission spectrum (Figure~\ref{fig:emission_spectrum}). We found that this spectrum is consistent with the published spectrum from \citet{Zieba2026} using the first three eclipse observations, which we plot for comparison.

\subsection{Stellar Spectrum}
In order to predict the wavelength-dependent secondary eclipse depth for a given planetary surface model, we must divide the predicted planetary emission spectrum by the stellar spectrum. We used  \texttt{Eureka!} following the process detailed in \citet{Zieba2026} to estimate the stellar flux in physical units during each of our eleven MIRI LRS observations and then calculated an averaged flux-calibrated stellar spectrum. This is the same approach taken in \citet{Zieba2026}, and it is motivated by the fact that there can be significant discrepancies between the measured and predicted infrared spectra of M dwarfs similar to LHS 3844~b \citep{Fauchez2025}.
 
In Stage 1, the only change we made was to set \texttt{skip\_emicorr} to True. In Stage 2, we set \texttt{skip\_photom} to False. We also increased the target and background aperture sizes to 18 and 19, respectively, to ensure that we captured all of the stellar flux. We only extracted the stellar flux from observations taken during the secondary eclipse (i.e., when the planet was behind the star, between 2nd and 3rd points of contact, see Figure~\ref{fig:white_light_curves}) to ensure that the planet's thermal emission did not contaminate the stellar spectrum. We checked for variability in the total stellar flux between observational epochs by integrating each spectrum across the \SIrange{5.06}{12.368}{\micro\meter}. We found that the observation taken on June 19 2023 (GO 1846 \#1) has the lowest flux at \SI{9.6577e4}{\watt/\meter^2} and the observation taken on October 3 2025 (GO 7953 \#7) has the highest flux at \SI{9.8954e4}{\watt/\meter^2}, respectively. This 2.4\% difference is comparable to the few-percent uncertainty in JWST's absolute flux calibration \citep{Law2025}. We also confirmed that using either the lowest or highest observed stellar spectrum in our calculations resulted in a negligible change in the predicted secondary eclipse depths. We therefore took the weighted mean of the 11 spectra and adopted this as the empirical stellar spectrum for our model fits.

In our fits, it is also sometimes useful to calculate a predicted dayside equilibrium temperature using a stellar spectrum that spans the full optical to infrared wavelength range. This is required in order to constrain the planetary albedo. We do not have an empirically measured stellar spectrum spanning the required wavelengths, so we instead used a SPHINX stellar spectral model for this purpose. We calculated this model using updated values for the stellar effective temperature $T_\mathrm{eff} = 3080\pm50$~\unit{\kelvin} and metallicity $\mathrm{[Fe/H]} = 0.22\pm0.10$~\unit{dex} from \citet{Nagel2026}. We find that the SPHINX stellar spectrum differs from our empirically measured mid-infrared JWST spectrum by up to 15\%, similar to what was presented in \citet{Zieba2026}. 

\section{Blackbody Fit and Feature Significance} \label{sec:results}
\subsection{Blackbody Fits}\label{sec:blackbody}
In order to determine the extent to which the observed spectrum is consistent with a blackbody function, we converted our wavelength-dependent secondary eclipse measurements into wavelength-dependent brightness temperatures (Fig. \ref{fig:emission_spectrum}). We found that LHS 3844~b has an effectively constant brightness temperature from 5~$\mu$m to 10~$\mu$m, with a dip longward of 10~$\mu$m. We fit the full spectrum with a two-parameter model where surface temperature and $R_p/R_*$ were allowed to vary in the fit in order to incorporate relevant uncertainties in the latter. We placed a uniform prior of $\mathcal{U}(900,1200)$~\unit{\kelvin} on the blackbody temperature and a Gaussian prior of $\mathcal{N}(0.6262, 0.00073)$ on $R_p/R_*$ based on the value reported in \citet{Nagel2026}. This resulted in a best-fit blackbody temperature of $T=997\pm11$~\unit{\kelvin} (log~$Z = 102.7$, $\chi^2=13.2$). This temperature is in good agreement ($0.3\sigma$) with the best-fit blackbody temperature of of $1000^{+15}_{-14}$~\unit{\kelvin} reported in \citet{Zieba2026}. We used this fitted brightness temperature to calculate an updated estimate of the brightness temperature ratio $\mathcal{R}_\mathrm{spec} = 0.95\pm0.03$, which is defined as the ratio between the measured dayside brightness temperature and the maximum predicted brightness temperature for a planet with an albedo of zero and no energy redistribution \citep[the latter condition, translating into $\mathcal{R}_\mathrm{spec} = 1$, is the default expectation for an airless body;][]{Coy2025}. This value remains close to unity, favoring a low dayside albedo, $A_B = 0.16\pm0.07$, and a low day-night redistribution efficiency for this planet. 

Although the emission spectrum is featureless shortward of $\sim10$~\unit{\micro\meter}, there appears to be a dip in brightness temperature between $10-12$~\unit{\micro\meter}. \cite{Zieba2026} also tentatively noted the presence of a similar feature in their analysis, but concluded that it was not detected at a statistically significant level in their three observations. If the surface of the planet is dominated by silicates, this feature might be an Si-O stretching feature \citep[the strongest diagnostic silicate spectral feature;][]{Salisbury1993, Hu2012, Paragas2025}. To determine the feature significance, we incorporated a Gaussian feature model into our blackbody fit, which now has three additional free parameters: the amplitude, wavelength center, and width ($\sigma$) of the Gaussian function. We placed priors on the Gaussian parameters based on the range of their best-fit values from fitting the Si-O stretching features of all of the solid samples in \citet{Paragas2025}: $\mathcal{U}(0,300)$~ppm for the amplitude, $\mathcal{U}(9, 13)$~\unit{\micro\meter} for the wavelength center, and $\mathcal{U}(0, 2.7)$~\unit{\micro\meter} for the width. We used Nested Sampling with \texttt{dynesty} \citep{Speagle2020, Koposov2025} to sample the posterior distributions of the free parameters and to estimate the Bayesian evidence $Z$. 

We found that the blackbody and Gaussian fit (log~$Z=104.9$) resulted in a global best-fit blackbody temperature of $991^{+16}_{-4}$~\unit{\kelvin}, amplitude of $204^{+35}_{-108}$~ppm, wavelength center of $11.4^{+1.2}_{-0.1}$~\unit{\micro\meter}, and width of $0.6^{+0.9}_{-0.1}$. The amplitude of the Gaussian feature is greater than zero at $1.9\sigma$ significance, and the Bayes factor of $B_m=9.4$ indicates substantial (almost strong) evidence in favor of the model with a Gaussian feature over the pure blackbody model \citep[see Table 1 of][]{Thorngren2026}. 

We fit the same blackbody and Gaussian model to the averaged emission spectrum assuming the best-fit NIRSpec phase curve shape (see Section~\ref{sec:lc_modeling}, Zieba et al., in preparation), and retrieved a slightly higher temperature of $1012^{+11}_{-10}$~\unit{\kelvin} and larger amplitude of $266^{+72}_{-76}$~ppm, which are $1.1\sigma$ and $0.5\sigma$ consistent, respectively, with our results assuming a simple sinusoidal phase curve model with zero nightside emission. For this case, the fitted Gaussian amplitude for the candidate feature is detected at higher significance ($3.5\sigma$), while the best-fit wavelength center and width are both $0.3\sigma$ consistent with the previous fit. In this fit, the data prefer the Gaussian model over a blackbody-only model with a Bayes factor $B_m=218$, indicating decisive evidence in favor of the feature. We note that the phase curve model we used in this fit is still preliminary and is based on unpublished data at shorter wavelengths than those spanned by our observations, so we conservatively adopt the simpler physically motivated sinusoidal phase curve model in all of our subsequent analyses moving forward. We note that a Gaussian shape is not a very good approximation of most of the complex and sometimes asymmetric Si-O stretching features in the \citet{Paragas2025} spectral library. We therefore refer readers to Section~\ref{sec:surface_composition} below, where we use a more realistic surface model to fit the feature.

\subsection{Can the Feature Be Explained by Residual Ramp Effects?}
\begin{figure}
    \centering
    \includegraphics[width=\linewidth]{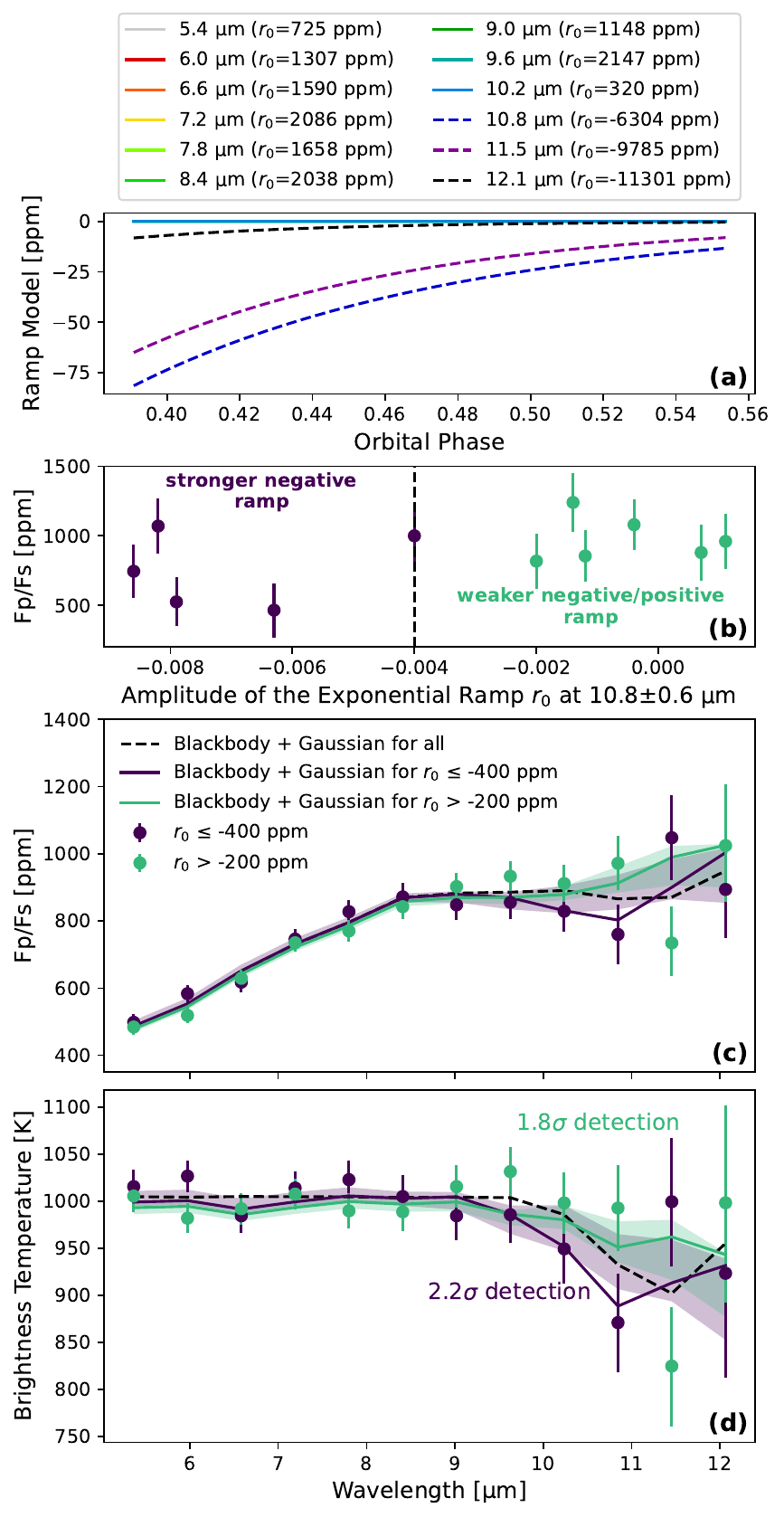}
    \caption{(a): The ramp models for all spectral channels for GO 7953 \#5 (October 1 2025) showing the change in ramp shape for channels in the shadowed region. (b): The planet-to-star flux ratios of each observation as a function of the ramp amplitude $r_0$ in the \SI{10.8\pm0.6}{\micro\meter} bin. (c): The averaged planet-to-star flux spectra for the two groups. (d): The same spectra converted to brightness temperature. (c) and (d) include the blackbody and Gaussian feature model (free temperature, amplitude, center, and width) fit to all 11 observations and each subgroup. The shaded regions correspond to the $1\sigma$ range of the best-fit models. The feature amplitude is detected to $2.2\sigma$ for the stronger ramp amplitude group and $1.8\sigma$ for the weaker ramp amplitude group.}
    \label{fig:ramp_effects}
\end{figure}

In the previous section, we evaluated the statistical significance of the observed feature assuming independent Gaussian distributed uncertainties for our wavelength-dependent secondary eclipse depths. However, previous studies have shown that MIRI LRS measurements frequently exhibit a sudden change in ramp shape for bins longward of \SI{10.5}{\micro\meter}, known as the shadowed region effect \citep{Bell2022}. Although there are not any obvious covariances between our retrieved eclipse depths and the exponential ramp coefficients in our fit, previous studies \citep[e.g.,][]{Agol2010} have shown that exponential ramps can subtly bias the retrieved eclipse depths. Although the underlying cause of the change in ramp shape at longer wavelengths is still under investigation, it is thought to be caused by uneven illumination of the subarray in the preceding observation. Because it depends on the previous illumination history of the array, it is not consistently present in all MIRI LRS observations. 

We examined our observations and found that there is a change in ramp shape longward of \SI{10}{\micro\meter} in several visits. An example of this behavior is shown in panel (a) of Figure~\ref{fig:ramp_effects}, in which the spectral channels shortward of \SI{10}{\micro\meter} have positive ramp amplitudes while the ones longward of \SI{10}{\micro\meter} have negative ramp amplitudes that invert the shape of the ramps from decreasing to increasing with time. We investigated whether this change in ramp shape might produce the long wavelength feature in our emission spectrum by plotting the amplitude of the exponential ramp $r_0$ and corresponding secondary eclipse depth in the \SI{10.8}{\micro\meter} bin for each of our 11 individual observations (Figure~\ref{fig:ramp_effects} (b)). We then subdivided the data into a set of five observations with stronger ramps (2023/06/19 GO 1846 \#1, 2024/05/04 GO 1846 \#3, 2025/06/16 GO 7953 \#2, 2025/10/01 GO 7953 \#5, and 2025/10/16 GO 7953 \#8) and six observations with weaker ramps, and calculated an averaged emission spectrum for each group (bottom two panels of Figure~\ref{fig:ramp_effects}). We then jointly fit the two emission spectra with the same blackbody and Gaussian feature model discussed in Section~\ref{sec:blackbody}, where we allowed the temperature and amplitude to vary between the two spectra but assumed a shared wavelength center and feature width.

We retrieved an amplitude of $210^{+34}_{-97}$~ppm ($2.2\sigma$) for the spectrum generated from the observations with more negative ramp amplitudes (corresponding to increasing flux with time) and 
an amplitude of $79^{+112}_{-45}$~ppm ($1.8\sigma$) for the spectrum generated from the observations with shallower ramp amplitudes. As expected, our retrieved surface temperatures, wavelength center, and feature width were all consistent with our previous fit to the full data set. Although the observations with more negative ramps appear to have a larger feature amplitude, the two retrieved amplitudes only differ by $0.4\sigma$. Additionally, we calculated the Pearson correlation coefficient between the secondary eclipse depths and amplitudes of the exponential ramps in the \SI{10.8}{\micro\meter} bin, and found a value of 0.47 with a 95\% confidence interval range of -0.18 to 0.83, indicating a relatively weak correlation. 

Next, we created another averaged spectrum, excluding the three observations with the strongest shadowed region effects (i.e., the three observations with the most negative $r_0$ in panel (b) of Figure~\ref{fig:ramp_effects}). We fit this spectrum with the same blackbody and Gaussian model, and retrieved an amplitude of $315^{+48}_{-57}$ ($5.5\sigma$), suggesting that the observations that are the most affected by the shadowed region may actually bias the fitted feature amplitude to lower values, weakening its detection significance. We conclude that the current data are too noisy to diagnose whether the observed feature is influenced by the exponential ramp shape.

\section{Bare Rock Surface Models}
\subsection{Surface Spectral Libraries}
If the tentative feature between $10-12$~\unit{\micro\meter} is astrophysical rather than instrumental in nature, we can identify representative surface compositions that might provide a good match to its measured shape. We modeled the emission spectrum of LHS 3844 b following the approaches in \citet{Hammond2025} and \citet{Paragas2025}. Although we initially considered models based on the surface types from \citet{Hu2012} and \citet{Paragas2025}, we found that none were a good match to our data. We therefore expanded our search into spectral libraries and used the bidirectional reflectance measurements from the RELAB database \citep{Milliken2021} to explore a wider variety of potential surface compositions. We followed the steps outlined in \citet{Hammond2025} to derive the single scattering albedo $\omega$ and the directional-hemispheric reflectance $r_\mathrm{dh}$ for a given material. We then used those values to calculate the hemispheric emissivity $\epsilon_h$ assuming isotropic scattering. This is an important caveat for our model predictions: most surfaces are not isotropic scatterers, but measurements obtained at a single scattering angle, such as those provided in RELAB, do not independently constrain the full angular scattering behavior of the samples needed to convert them into disk-integrated planetary spectra \citep[e.g.,][]{Pilorget16,Gkouvelis25_RNAAS}. However, these simplifying assumptions are sufficient for our purpose here, which is to identify simple end-member surface compositions that might have features in the desired wavelength range.  

After calculating the hemispheric emissivity, we followed the steps in \citet{Paragas2025} to calculate the outgoing planetary flux. For each surface type, we also derived the 1D correction factor $f$. This factor is calculated by comparing our 1D surface model to the output of a 3D surface model from \citep{Hu2012} that calculates the equilibrium temperature at every point on the planet. 

\subsection{Surface Composition}\label{sec:surface_composition} 
In principle we can calculate a predicted equilibrium surface temperature and corresponding dayside emission spectrum for each of these materials using their measured reflectances \citep[e.g.,][]{Hu2012, Paragas2025, Zieba2026}. However, if the planetary surface includes a mixture of several materials, this can result in a wide range of optical albedos while preserving infrared spectral features from individual mineral constituents. Similarly, space weathering from micrometeoroid bombardment and/or excess irradiation \citep[e.g., solar wind and cosmic rays;][]{Pieters2016, Lyu2024} can systematically darken older surfaces on airless planetary bodies, increasing their corresponding equilibrium dayside temperatures. Particle size also exerts substantial control on albedo \citep{Pieters1983}. We therefore allowed the dayside temperature to vary as a free parameter in our fit. 

The Christiansen feature (CF) is  an emissivity maximum in the spectrum. In silicates, it is located at the beginning of the Si-O stretching feature, and its wavelength position is correlated with the \ce{SiO2} wt\% of the material \citep{Conel1969}. This effect can be measured in spectrally coarse resolution data of planetary surfaces \citep{Glotch2010}. Specifically, if particle size and surface temperature gradient are constant, the CF peak occurs at longer wavelengths for lower \ce{SiO2} wt\% and at shorter wavelengths for greater \ce{SiO2} wt\%. Particle size, particle packing, and surface temperature gradients also influence the position \citep{Logan1973, DonaldsonHanna2012}. In order to match the relatively long wavelength of the candidate CF observed in our data, near \SI{10}{\micro\meter}, as well as the brightness temperature minimum near \SI{11.5}{\micro\meter} we selected materials with features that occur at similar wavelengths, including olivine iron and magnesium endmembers fayalite and forsterite, respectively, spinels \citep[found in primitive Solar System materials and some igneous rocks, e.g.,][]{Burbine1992, Sunshine2008}, and SiC \citep[a possible exotic composition for rocky exoplanets;][and references therein]{Guimond2024} from the RELAB database. We selected these materials to demonstrate that there are multiple candidate surface compositions that may be able to match the long-wavelength feature in our observations. 

For each surface type considered, we identified samples in RELAB with a range of grain sizes and/or packing methods. We considered two olivines, fayalite and forsterite. We found that the fayalite sample (ID: DD-MDD-098-P; 0-45~\unit{\micro\meter} grains that have been pressed into a pellet) provided the best overall fit ($\chi^2 = 3.6$ for 12 data points and two free parameters, temperature and the ratio of the planet to stellar radii) with a retrieved temperature of $1020^{+12}_{-11}$~\unit{\kelvin}. This $\chi^2$ is significantly lower than the $\chi^2$ value of the blackbody only fit of 12.9, indicating an improvement in fit quality. Although it is also lower than $\chi^2_\nu \approx1$, where $\chi^2_\nu = \chi^2/\nu$ and $\nu = n-m$ with $n$ being the number of data points and $m$ being the number of free parameters, this is not surprising given the large number of available materials in RELAB, which collectively span a wide range of spectral feature depths, widths, and locations. 

This specific fayalite powder sample was pressed into a pellet, making its spectrum more closely resemble that of a solid slab than a loosely packed powder sample \citep{Salisbury1987}, effectively eliminating the transparency feature \citep[e.g.,][]{Paragas2025}. To determine whether or not we see evidence for the transparency feature, we also fit a model using the fayalite sample (ID: DH-MBW-006) with loosely packed 0-30~\unit{\micro\meter} grains. This sample was disfavored ($\chi^2 = 40.8$, ${1089\pm13}$~\unit{\kelvin}) by the data due to the combination of the rising slope between \SIrange{7}{10}{\micro\meter} and the shallower Si-O stretching feature caused by the transparency feature (see the bottom panel of Figure~\ref{fig:emission_spectrum}). None of the forsterite samples were able to match the observed location of the candidate CF in our observations, resulting in a lower quality fit ($\chi^2 = 39.8$, ${1066\pm12}$~\unit{\kelvin}) for our preferred sample (ID: DH-MBW-009; 64-120~\unit{\micro\meter} grains). We note that the available forsterite samples were also loosely packed powders, and therefore displayed a transparency feature that was also disfavored by our data.

Spinel has been found in early-formed materials, such as calcium aluminum inclusions \citep[e.g.,][]{Sunshine2008}, $>4$~Ga old asteroids \citep[e.g.,][]{Burbine1992}, and in assemblages with mafic to ultramafic rocks on the Moon \citep[e.g.,][]{Pieters2011}. It also has an emissivity minimum feature at wavelengths similar to the one in our data. However, the spinels available in RELAB were limited to loosely packed powder samples with very strong transparency features. We selected one ($\chi^2 = 99.2$; ID: SP-CMP-077-C; 45-75~\unit{\micro\meter} grains) to show as an example in Figure~\ref{fig:emission_spectrum}. We found that this spinel exhibits a feature around \SI{11}{\micro\meter} that is a potential compositional match, but the loosely packed particle texture characteristic of this particular RELAB sample is disfavored due to the presence of a strong transparency feature that manifests in the spectrum as a steep slope between 5 and 10.3~\unit{\micro\meter}. In the future, additional lab measurements of larger grained  and/or pressed spinel samples could provide an improved fit to our observations. 

We also found that silicon carbide, SiC, surface models provided a reasonable match to our data. SiC has been suggested as a possible exotic composition for rocky exoplanets that have formed in high C/O environments \citep{Hakim2018, Hakim2019, Miozzi2018, Allen-Sutter2020, Guimond2024}, and observed as dust around carbon-rich stars \citep{Speck2009, Sloan2014}. There was only one SiC sample (ID: MM-PTB-001; a silicon carbide slab) available in RELAB with sufficient mid-IR wavelength coverage. Although this sample has an absorption feature at a wavelength similar to that observed in our data, the SiC feature in our best-fit surface model is far too deep ($\chi^2=313.1$). We therefore considered a linear mixture of the SiC sample and the K1919 basalt powder from \citet{Paragas2025}, which we chose, not for geological plausibility, but solely because it is a dark and relatively featureless material. We found that a mixture of 63\% SiC and 37\% K1919 basalt powder provides a better fit to the data ($\chi^2=5.6$, \SI{1087\pm14}{\kelvin}). The best-fit models for all surface types considered are shown in Figure~\ref{fig:emission_spectrum}.

Both laboratory and theoretical studies have shown that the depths of surface spectral features seen in emission are sensitive to the unknown near-surface temperature gradient \citep[e.g.,][]{DonaldsonHanna2012,Paragas2025,Lyu26}. We therefore repeated our fit for the fayalite surface model, but this time we allowed the depth of the Si-O stretching feature to vary as a free parameter in the fit. We also allowed the location of the feature to vary in the fit, allowing us to obtain a corresponding constraint on the location of the CF. For this test, we modeled the surface as a blackbody with the fayalite Si-O stretching feature superimposed onto it. We used Nested Sampling with \texttt{dynesty} to sample the posterior distributions of the blackbody temperature, the depth of the Si-O stretching feature, and the CF wavelength (from now on called the three parameter model). We set broad prior ranges of $\mathcal{U}(900,1200)$~\unit{\kelvin} for the blackbody temperature, $\mathcal{U}(0,300)$~ppm for the Si-O stretching feature depth, and $\mathcal{U}(5,12)$~\unit{\micro\meter}, and a Gaussian prior of $\mathcal{N}(0.6262, 0.00073)$ on $R_p/R_s$ based on \citet{Nagel2026}. For this three parameter model, assuming literature values for system parameters, we retrieved a temperature of \SI{996+-10}{\kelvin}, an Si-O stretching feature depth of $188^{+61}_{-67}$~ppm, and a CF wavelength of $10.0^{+0.4}_{-0.3}$~\unit{\micro\meter}. 

\subsection{Constraints on Surface Albedo from Energy Balance}
For each of the surface types considered above, 
the planet's measured dayside flux implicitly constrains its optical and near-infrared albedo. Composition, grain size, packing and space weathering all influence albedo, and its value therefore provides an independent constraint on the planet's dayside surface properties. We quantified this constraint by creating a grid of 3D equilibrium surface models for each surface type using the model framework from \cite{Hu2012} with varying Bond albedos $A_B$. At each grid point, we rescaled the incident flux absorbed by the surface using a scale factor of ($1-A_{B,new})/(1-A_{B,orig}$) where $A_{B,new}$ is the new assumed Bond albedo and $A_{B,orig}$ is the original Bond albedo for that surface type. This ensured that we correctly accounted for the changing 3D dayside temperature distribution as we varied $A_B$. We also accounted for uncertainties in $T_*$ and $a/R_*$ by randomly sampling from Gaussian distributions set by values in \citet{Nagel2026} of $\mathcal{N}(3080, 50)$~\unit{\kelvin} and $\mathcal{N}(7.122, 0.078)$, respectively. For the fayalite sample, we retrieved a Bond albedo of $0.18\pm0.10$. This is in good agreement with its true Bond albedo of $0.17$, which we calculated using our SPHINX stellar spectrum and the measured optical to near-infrared reflectivity of the fayalite sample. 

For the SiC and basalt mixture, we retrieved a Bond albedo of $0.08\pm0.05$, which is $3\sigma$ less than its true Bond albedo of 0.23. This surface has lower emissivities between $5-10$ \unit{\micro\meter} than the fayalite sample, and therefore requires higher surface temperatures to match our measured flux. This lower albedo might plausibly be explained by space weathering \citep[e.g.,][]{Lyu2024}. However, there may also be other grain sizes, packing, or compositional mixtures including this material that can match the observed long-wavelength feature while simultaneously allowing for a lower dayside temperature. Angle-dependent effects, such as the opposition surge, that we neglected in our simplified modeling framework could also result in a higher predicted dayside flux for the same underlying surface composition and temperature \citep[e.g.,][]{Gkouvelis25_PSJ}. 

\section{Conclusions} \label{sec:discussion_and_conclusions}
We present an updated MIRI LRS eclipse spectrum of LHS 3844 b by combining 11 (8 new and 3 archival) eclipse observations. The data are well-matched by a \SI{996\pm5}{\kelvin} blackbody shortward of \SI{10}{\micro\meter}, and display a tentatively identified feature ($1.9-3.5\sigma$) between $10-12$~\unit{\micro\meter}. We explore whether this feature might be produced by the ramp shape changes associated with the shadowed region, and are unable to reach a definitive conclusion. If this feature is astrophysical, its depth and location are consistent with ultramafic materials, such as olivine endmembers fayalite (iron) and forsterite (magnesium), spinels, or silicon carbide SiC. The data disfavor the presence of a strong transparency feature, indicating that the surface is unlikely to be predominately comprised of small loosely packed grains. Assuming a blackbody, the inferred Bond albedo is $A_B = 0.16\pm0.07$. In the future, we expect that expanded modeling and/or laboratory measurements of a wider range of geologically plausible mixtures may identify additional models that can provide a good match to these observations. The simple model framework that we utilized here could also be improved by a more detailed consideration of angle-dependent scattering effects \citep[e.g.,][]{Gkouvelis25_RNAAS,Gkouvelis25_PSJ}, near-surface temperature gradients \citep[e.g.,][]{Lyu26}, and temperature-dependent changes in the wavelength locations of spectral features \citep[e.g.,][]{Fortin24,Paragas2025}, as well as other related phenomena.

Our observations demonstrate that JWST can achieve the mid-infrared sensitivity required to spectroscopically constrain the surface properties of rocky exoplanets. In the future, near-infrared observations of LHS 3844~b's emission spectrum with NIRSpec (GO 4008, PI: S. Zieba and GO 7953, PI: K. Paragas) can be used to obtain improved constraints on the strength of the transparency feature, which is strongest between $3-8$~\unit{\micro\meter}. This feature is caused by multiple scattering and provides a relatively composition-insensitive diagnostic of the surface particle size and texture, which is related to the surface age \citep{Paragas2025}. At the same time, new MIRI LRS observations are needed in order to investigate the possible influence of the instrumental ramp on the observed $10-12$~\unit{\micro\meter} feature and to collect data of sufficient quality to differentiate between compositional models. The new MIRI LRS subarrays (SLITLESSPRISM\_IP and SLITLESSPRISM\_IPS), which was added in Cycle 6, will mitigate wavelength-dependent changes in the ramp shape by shifting the location of the subarray away from the shadowed region. Future observers might also consider using the ``pre-flash" strategy first developed for mid-infrared Spitzer timeseries photometry \citep[e.g.,][]{Deming2009, Knutson2011}. This approach exposes the detector to a bright, diffuse source immediately before the science observations, which effectively eliminates the ramp behavior in most observations.

\begin{acknowledgments}
We thank Geoffrey Blake for useful discussions about silicon carbide. This work is based on observations made with the NASA/ESA/CSA James Webb Space Telescope. The data were obtained from the Mikulski Archive for Space Telescopes at the Space Telescope Science Institute, which is operated by the Association of Universities for Research in Astronomy, Inc., under NASA contract NAS 5-03127 for JWST. These observations are associated with GO program 7953. Support for US investigators in program 7593 was provided by NASA through a grant from the Space Telescope Science Institute, which is operated by the Association of Universities for Research in Astronomy, Inc., under NASA contract NAS 5-03127. All of the data presented in this article were obtained from the Mikulski Archive for Space Telescopes (MAST) at the Space Telescope Science Institute. The specific observations analyzed can be accessed via \dataset[doi: 10.17909/gppv-t130]{https://doi.org/10.17909/gppv-t130} and \dataset[doi: 10.17909/94sq-tr83]{https://doi.org/10.17909/94sq-tr83}. The models used in this study were developed at the Jet Propulsion Laboratory and the California Institute of Technology under a contract with the National Aeronautics and Space Administration and funded through the President’s and Director’s Research \& Development Fund Program. S.Z. was supported by NASA through the NASA Hubble Fellowship grant \#HST-HF2-51570.001-A awarded by the Space Telescope Science Institute, which is operated by the Association of Universities for Research in Astronomy, Incorporated, under NASA contract NAS5-26555. 
\end{acknowledgments}

\facilities{JWST (MIRI LRS)}

\software{\texttt{astropy} \citep{AstropyCollaboration2013, AstropyCollaboration2018, AstropyCollaboration2022},  
          \texttt{batman} \citep{Kreidberg2015},
          \texttt{dynesty} \citep{Speagle2020, Koposov2025},
          \texttt{emcee} \citep{Foreman-Mackey2013},
          \texttt{Eureka!} \citep{Bell2022}
          }

\appendix
\section{Individual Visits}
\begin{figure}
    \centering
    \includegraphics[width=\linewidth]{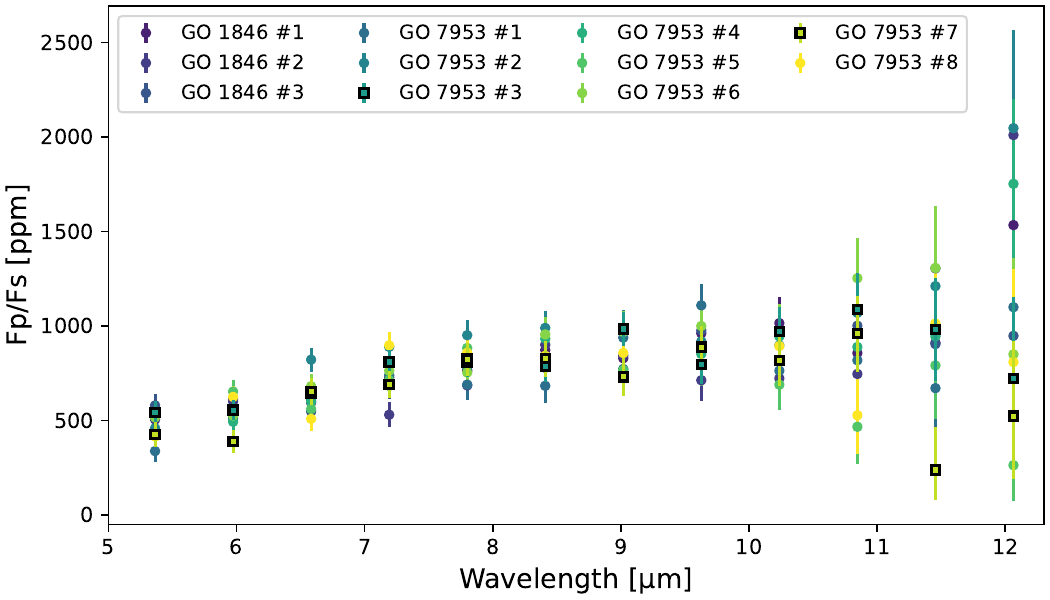}
    \caption{The individual emission spectra of each observation. The two sets of square markers outlined in black are the two visits with noted excess noise in Section~\ref{sec:methods}.}
    \label{fig:individual_spectra}
\end{figure}

As discussed in Section~\ref{sec:methods}, we constructed the average emission spectrum shown in Figure~\ref{fig:emission_spectrum} by taking the inverse-weighted average of the 11 individual eclipse depths in each bandpass. The eclipse depth spectrum for each individual observation is shown in Figure~\ref{fig:individual_spectra}. We highlight the two visits with the most correlated noise, GO 7953 \#3 from September 29 2025 and GO 7953 \#7 from October 3 2025, by marking them as squares outlined in black. They appear to be consistent with the other visits. We note that for GO 7953 \#7, the two longest wavelength eclipse depths are relatively low, but their uncertainties and the overall scatter among all visits is larger in these bins.

\section{Stellar Spectra}
\begin{figure}
    \centering
    \includegraphics[width=\linewidth]{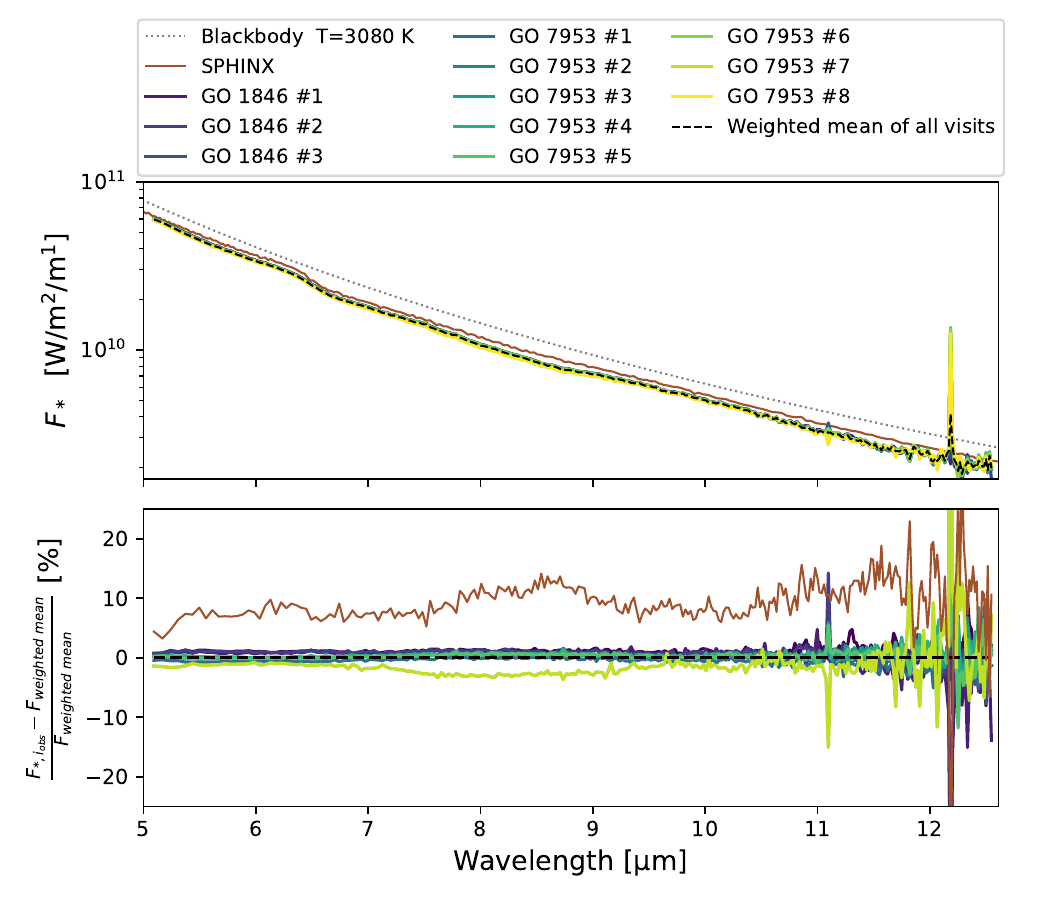}
    \caption{Top: The spectra of a blackbody (3080~\unit{\kelvin}, the SPHINX model, each individual visit and the weighted mean of all observations. Bottom: The percent difference of each spectra relative to the weighted mean of all visits.}
    \label{fig:stellar_spectra}
\end{figure}

We show the absolute flux calibrated stellar spectra from each visit in Figure~\ref{fig:stellar_spectra} for further comparison aside from the brief discussion in Section~\ref{sec:methods} on the stellar spectrum. Most of the spectra overlap across the wavelength range except for GO 7953 \#7 and \#8, whose spectra are lower between \SI{\sim6.8}{\micro\meter} and \SI{\sim10.5}{\micro\meter}. Additionally, we note a similar behavior, as first mentioned in \citet{Zieba2026}, where the SPHINX spectrum overestimates the flux of the star by $\sim15\%$. This further highlights the important of using absolute flux calibrated spectra for interpreting future datasets \citep{Fauchez2025}.

\bibliography{sample701}{}
\bibliographystyle{aasjournalv7}

\end{document}